\documentclass[sigconf]{acmart}

\usepackage{booktabs} 

\begin{document}

\copyrightyear{2018}
\acmYear{2018}
\setcopyright{acmcopyright}
\acmConference[RecSys '18]{Twelfth ACM Conference on Recommender Systems}{October 2--7, 2018}{Vancouver, BC, Canada}
\acmPrice{15.00}
\acmDOI{10.1145/3240323.3240397}
\acmISBN{978-1-4503-5901-6/18/10} 

\title[Attentive Neural Arch. Incorporating Song Features For Music Reco.]{Attentive Neural Architecture Incorporating Song Features For Music Recommendation}

\author{Noveen Sachdeva}
\affiliation{%
  \institution{International Institute of Information Technology}
  \city{Hyderabad, India} 
}
\email{noveen.sachdeva@research.iiit.ac.in}

\author{Kartik Gupta}
\authornote{Noveen Sachdeva and Kartik Gupta had equal contribution towards the research work demonstrated in the paper.}
\affiliation{%
  \institution{International Institute of Information Technology}
  \city{Hyderabad, India} 
}
\email{kartik.gupta@research.iiit.ac.in}

\author{Vikram Pudi}
\affiliation{%
  \institution{International Institute of Information Technology}
  \city{Hyderabad, India} 
}
\email{vikram@iiit.ac.in}

\renewcommand{\shortauthors}{N. Sachdeva et al.}

\begin{abstract}
Recommender Systems are an integral part of music sharing platforms. Often the aim of these systems is to increase the time, the user spends on the platform and hence having a high commercial value. The systems which aim at increasing the average time a user spends on the platform often need to recommend songs which the user might want to listen to next at each point in time. This is different from recommendation systems which try to predict the item which might be of interest to the user at some point in the user lifetime but not necessarily in the very near future. Prediction of next song the user might like requires some kind of modeling of the user interests at the given point of time. Attentive neural networks have been exploiting the sequence in which the items were selected by the user to model the implicit short-term interests of the user for the task of next item prediction, however we feel that features of the songs occurring in the sequence could also convey some important information about the short-term user interest which only the items cannot. In this direction we propose a novel attentive neural architecture which in addition to the sequence of items selected by the user, uses the features of these items to better learn the user short-term preferences and recommend next song to the user.
\end{abstract}

%

\ccsdesc[500]{Information systems~Recommender systems}
\ccsdesc[300]{Information systems~Content ranking}

\keywords{Recommender Systems; Short Term Interest}

\maketitle

\section{Introduction}
There has recently been an intense focus on recommendation systems by the Information Retrieval community because of their commercial experience and the ability to provide a better experience to the user while interacting with a large database of items. Often there are a very large number of items in the database that might be of interest to the user, to the extent that the user might not even know they exist. Hence, they need to be presented to the user as a recommendation. To give an example, for websites which sell different kinds of products and have a huge catalog, users might feel better if they didn't have to browse for the items they might like and were rather recommended by the system, saving time and effort of the user, thus creating a pleasant experience.

The content of the item chosen by the user is often an indication of the items that might be of interest to the user. In the case of music, this might not always be constant and might change with time. In a recent work Gupta~\cite{own_work}, tries to model the short-term preferences of the user for music recommendation. He uses Last.fm~\cite{last_fm} tags to find out song features important to the user instead of the content derived from the audio. Last.fm tags look promising in describing the contents of the song and also provide a lot more information about the song which could be very hard to derive either from the audio or the metadata of the song. We align to the claim that Last.fm could very well be used to model the song features which might be of interest to the user. However, the similarity function used by Gupta could be better learned and provide a better performance. Gupta also claims that it is the group of items that occur together which matter while recommendation and not the exact sequence in which they occur.

Towards this claim made by Gupta and finding a better similarity function, we apply Attentive Neural Networks to the problem of next item Prediction. Attentive neural networks indeed give different weights to each item in the sequence and the weights are not in order of the items. The third last item selected by the user could get more weight than the last item selected by the user and hence the choice of Attentive Neural Networks takes the claim into account. Also, we introduce a content attention component, which deals with the tags of the items, assuming these tags indeed can model the short-term interests of the user. This component takes the tags of the items selected by the user in the recent past.

\section{Related Works}

Recommender systems is a well-researched topic and a wide variety of systems have been developed and it is important that we cover some of them here to provide a context to the reader.

\subsection{Collaborative Filtering}\label{subsec:CF}
It exploits the $user-item$ interactions to find similar users based on the number of same items selected. A variant is item level collaborative filtering~\cite{cf2}, wherein two items selected by the same user are considered to be similar. 

There have been improvements to collaborative filtering such as matrix factorization~\cite{cf1} of the $user-item$ matrix into the $user$ feature matrix and the $item$ feature matrix. Further, there have been ranking algorithms such as Bayesian personalized ranking~\cite{bpr} to further provide better and personalized recommendation to users. 

\subsection{Content Based Recommendation}\label{subsec:CBR}
Content-based systems recommend items based on the similarity of content to the items already selected by the user~\cite{content1,content2}.
If the content of a song is similar to the ones the user likes, then that song is more probable to be recommended to the user. For example, there are systems which recommend songs based on the melody of the song ~\cite{content4}. Another example which also assumes that the tags can indeed be sufficient to model the features of the songs which might be of importance to the user is by Liang~\cite{content3} which generates a latent vector for each song based on the semantic tags and then applies collaborative filtering to provide a recommendation to the users.

\subsection{Sequence Based Recommendation}\label{subsec:SBR}
Recommendation can be modeled as a sequence prediction problem and the first attempt at it was by Brafman~\cite{seq3}. The initial attempts were based on simple models such as Markov chains and they have been further improved. One such improvement is having a personal Markov chain for each user~\cite{seq1}. With the popularity of recurrent neural networks, they have been applied~\cite{seq2} to the problem of next item prediction and have performed much better than the other systems.
With the success of Attentive neural networks in fields such as language and speech processing, they have been applied to recommender systems as well~\cite{attentive_1}. Our model applies attention to the sequence of items as well as the content of those items. Two context vectors are computed in the model independently, one which gives a context solely based on the items and the other which gives a context based only on the tags of the items. 

\subsection{Hybrid Recommender Systems}\label{subsec:hybrid}
Hybrid systems combine two or more techniques in order to provide better recommendations. Yoshii~\cite{hybrid1} proposed a system wherein the recommendations are based on the rating as well as the content, which are modeled based on the polyphonic timbres of the song. Hariri~\cite{hariri} applied topic modeling and models the sequence of songs heard by the user as a sequence of topics and then tries to predict the next topic and the next song in that topic. The transitions between topics are learned from a collection of playlists. Gupta~\cite{own_work} proposes a hybrid model which takes into account the different songs played together and the tags of the song. The approach is able to tell at any given point of time the features of the songs the user is interested in. Shobu~\cite{hybrid10} builds an interesting system which bases its recommendation on the transition of acoustic features over the songs. It tries to generate a sequence of songs over which the transition of acoustic features is smooth.

\section{Our Method}

\begin{figure*}[ht!]
 \centering
  \includegraphics[width=1\textwidth]{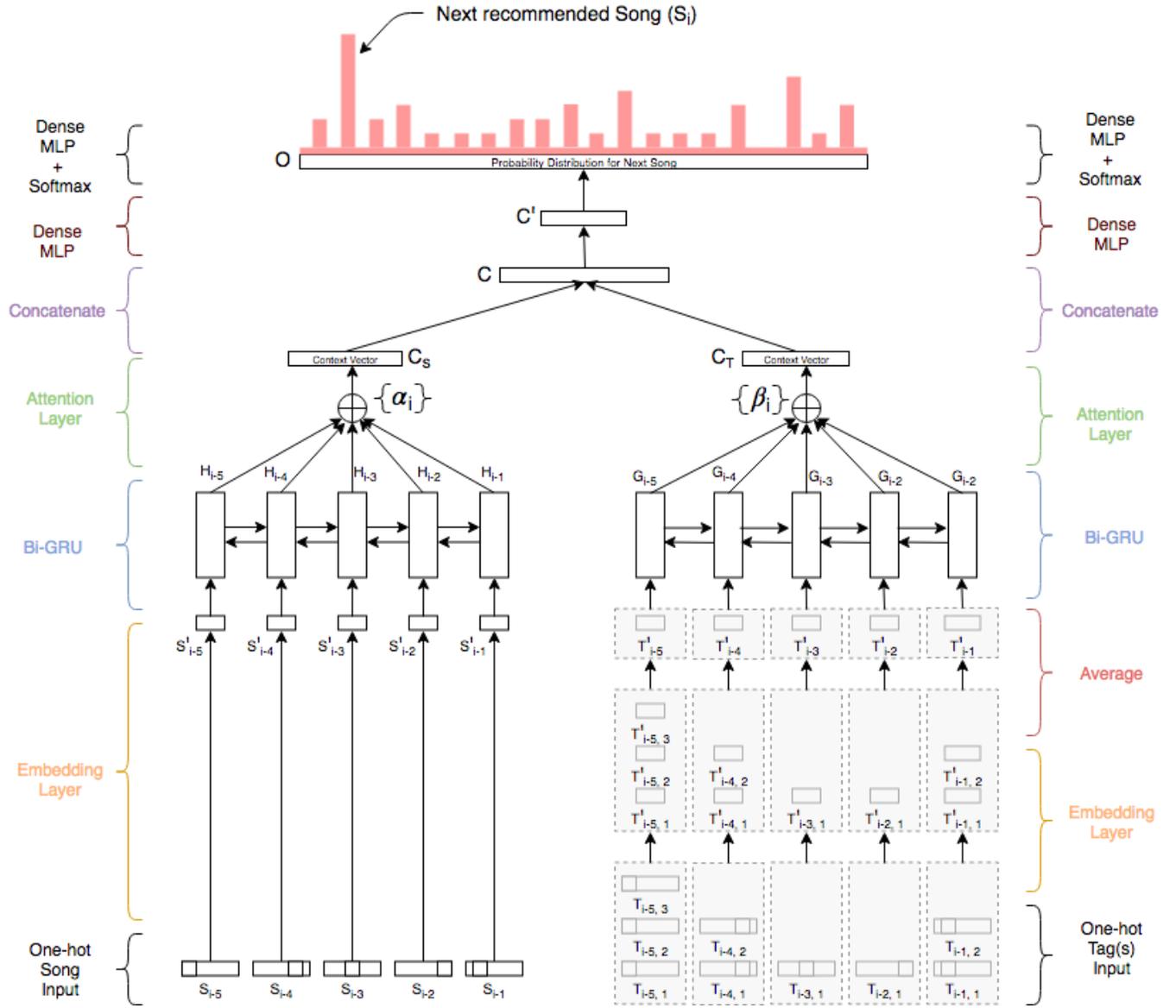}
 \caption{Attentive Neural Network Architecture for Next Song Prediction}
 \end{figure*}
We present an Attentive Neural Architecture to tackle the problem of next item prediction which has the ability to include tags of the items and models the short term user interests based on the features of the items as well as the items themselves. We now present the formal problem statement that we try to tackle in this paper.

\textbf{Predicting Next Song} \textit{Given the set of songs heard by the user in sequence $S_{s} = \{s_{1}, s_{2}, ..., s_{i-1}\}$} and the tag set for each song, $T_{i} = \{t_{i}^{1}, t_{i}^{2},..., t_{i}^{j} \}$, predict $s_{i}$. 

\subsection{Proposed Solution}
The architecture we propose is shown in figure 1. The output of the model are the probabilities of each item occurring next, given the items occurred in the user history ($P(s_{i} \mid s_{i-1}, s_{i-2}...s_{i-m})$). The first component receives as input the one hot encoding of the songs which occurred before the song to be predicted. The second component receives the one hot encoding of all the tags for the items occurring before the song to be predicted. The song-embedding layer maps the one hot representations of the songs to a vector space which are then fed to a Bi-GRU in the first component. Similarly, the tags for each song are also converted to their distributed representations using another embedding layer. For each song, the average of the distributed representations of all its tags is fed to a Bi-GRU in the second component. For both components, the hidden states are given as input to an attention layer where the attention-score or weight for each hidden state is computed. The output of the attention layer is the context vector which is the weighted sum (given by the attention layer) of the hidden states of the Bi-GRU. The context vectors coming from both components are concatenated and fed to a smaller dimension non-linear dense layer, using ReLU as the activation function. The output of this dense layer is then fed to another dense layer followed by a softmax operation, used to calculate probabilities over all songs modeling the next song. Below we present the equations for a better understanding of the model. Let $V= \{ v_{1}, v_{2} ... v_{\mid V \mid}\}$ be the set of all the songs.

\begin{equation}
    \begin{aligned}
    s'_{i} = E_{1} * s_{i}
    \end{aligned}
\end{equation}
where $s_{i}$ is the one hot representation of the song, $E_{1}\in R^{d*\mid I \mid}$ is the embedding layer, $d$ is the length of the embedded song vector and $\mid I \mid$ is the set of all songs.

\begin{equation}
    \begin{aligned}
    t^{' j}_{i} = E_{2} * t_{i}^{j}
    \end{aligned}
\end{equation}
where $t_{i}^{j}$ is the one hot representation of the $j^{th}$ tag of the $i^{th}$ song,  $E_{2}\in R^{d'*\mid T \mid}$ is the embedding layer, $d'$ is the length of the embedded tag vector and $\mid T \mid$ is the set of all tags.

\begin{equation}
    \begin{aligned}
    t^{'}_{i} = \frac{1}{n_{i}} \sum_{j=1}^{n_{i}} t^{' j}_{i}
    \end{aligned}
\end{equation}

where $n_{i}$ is the number of songs associated to the $i^{th}$ song, and $t'_{i}$ is the average of the embedding vector of all the tags associated to the $i^{th}$ song.

The hidden states of both Bi-GRUs, $H_{i}$ and $G_{i}$, which are fed to the attention layer are a mere concatenation of the two individual unidirectional hidden states: $\overset{\rightarrow}{h_{i}}, \overset{\leftarrow}{h_{i}}$ and $\overset{\rightarrow}{g_{i}}, \overset{\leftarrow}{g_{i}}$ respectively.

Both the attention layers output a context vector which is a weighted sum of all the hidden states. $C_{s}$ is the context vector computed from the song component of the model and $C_{t}$ is the context vector computed from the tag component of the model. 
\begin{equation}
    \begin{aligned}
    C_{s} = \sum_{j=i-1}^{i-m}\alpha_{j}H_{j}
    \end{aligned}
\end{equation}
\begin{equation}
    \begin{aligned}
C_{t} = \sum_{j=i-1}^{i-m}\beta_{j}G_{j}
\end{aligned}
\end{equation}

Both the context vectors, $C_{s}$ and $C_{t}$ are then concatenated resulting in a final context vector, $C$ which is then fed to a dense layer using the standard equations. 

\begin{equation}
    \begin{aligned}
    C' = ReLU(W_{1}C + b_{1})
    \end{aligned}
\end{equation}

$C'$ is nothing but a vector representation of $C$ in a smaller dimension vector space which significantly reduces the training time because the following dense layer has a huge dimension (Number of songs).
The final output is a dense layer of the size of the total number of songs followed by a softmax function which gives the probability of occurrence of each song given the user's history.
\begin{equation}
    \begin{aligned}
    O = W_{2}C' + b_{2}
    \end{aligned}
\end{equation}
\begin{equation}
    \begin{aligned}
    P(v_{l}=s_i \mid s_{i-1}, s_{i-2}, ..., s_{i-m}) = \frac{e^{v_{l}}}{\sum^{\mid V \mid}_{p=1}e^{v_{p}}}
    \end{aligned}
\end{equation}
 Negative log likelihood was used as the loss function and the optimization problem becomes:

\begin{equation}
    \begin{aligned}
    \underset{X, Y, W_{1}, W_{2}, b_1, b_2}{argmin} -\sum_{s''}\sum_{t}logP(v_{l}=s_i \mid s_{i-1}, s_{i-2}, ..., s_{i-m})
    \end{aligned}
\end{equation}

 where $s''$ is a user session in the dataset and $v_{l}$ is the actual song which occurs after the $m$ given songs. $X$ and $Y$ are the matrices consisting of song and tag embeddings respectively. We iterate over all the sessions in the datasets and all time steps in those sessions.

\section{Experiments}

\subsection{Dataset}
The dataset was a subset taken from the Last.fm dataset \cite{last_fm}. Each log in the dataset consisted of user id, song name, artist name and time stamp. We performed experiments on a subset consisting of 6-month histories of all the users and the tags for each song were retrieved using the Last.fm public API. The user histories were divided into sessions as done by Gupta~\cite{own_work}. The first 70 percent of the sessions for each user (in order of occurrence)  were put in the training set and the last 30 percent in the test set. Sessions having less than 5 songs were discarded.

\begin{table}
\begin{center}
 \begin{tabular}{|l|l|}
  \hline
  \textbf{Description} & \textbf{Value} \\
  \hline
  Total Logs  & 3553321 \\
  \hline
  Total Users & 759 \\
  \hline
  Total Sessions  & 110410 \\
  \hline
  Total Unique Songs & 386046 \\
  \hline
  Total Unique Tags & 487844 \\
  \hline
  Average Songs Per Session & 32.18 \\
  \hline
  Average logs per user & 4681.58 \\
  \hline
 \end{tabular}
 \end{center}
 \caption{Dataset Statistics}
 \label{tab:example}
 \end{table}

\begin{table}
 \begin{center}
 \begin{tabular}{|l|l|l|l|l|l|}
  \hline
  \textbf{Model} & \textbf{k=10} & \textbf{k=20} & \textbf{k=30} & \textbf{k=40} & \textbf{k=50}\\
  \hline
  POP  & 0.85 & 0.97 & 1.24 & 1.69 & 2.14\\
  \hline
  BPR-MF  & 7.34 & 8.13 & 8.56 & 8.98 & 9.27 \\
  \hline
  SSCF & 13.69 & 17.12 & 19.66 & 21.30 & 22.34 \\
  \hline
  RNN & 14.42 & 16.26 & 16.74 & 17.09 & 17.38 \\
  \hline
  SBRS & 19.15 & 26.14 & 28.83 & 30.35 & 31.40\\
  \hline
  SABR & 26.36 & 28.61 & 29.97 & 31.72 & 32.47\\
  \hline
  STABR & \textbf{28.95} & \textbf{30.85} & \textbf{31.90} & \textbf{32.65} & \textbf{34.26}\\
  \hline
  
 \end{tabular}
\end{center}
 \caption{Results}
 \label{tab:example2}
\end{table}

\subsection{Baselines}
The architecture is tested against the following baselines:
\begin{enumerate}
  \item POP: The most popular items in the training set are recommended to the users. 
  \item BPR-MF: A matrix factorization based model which ranks items for each user \cite{bpr} differently. The implementation by MyMediaLite was used with default parameters except for the number of features which was kept 100 for best results. We report the mean over 5 runs for this model.
  \item Session Based Collaborative Filtering(SSCF): This system instead of making a $user-item$ matrix makes a $session-item$ matrix and recommends items by finding similar sessions in the database to the active session based on the songs which have already occurred in the current session. The similar sessions were found based on the last 5 songs heard by the user and the results are reported based on 100 nearest sessions.
  \item RNN: In this method, the sequence of items occurring together is fed to a recurrent neural network trying to predict the next item at each timestep. All sequences in the train set are used to learn the model and to get the next recommendation, all the songs heard by the user until that point are fed to the network. We used the implementation provided by the authors of ~\cite{umap} based on mini batch stochastic gradient descent and we kept the batch size to 20, using the Categorical Cross Entropy loss function with a 100 hidden units for the RNN and a learning rate of 0.1. 
  \item Subsession Based Recommender System: This method was proposed by Gupta~\cite{own_work}. In this method, short-term user preferences are found using the tags of the songs the user heard. The user history is divided into small windows of constant preference and songs are found based on the similar window in the training set to the active window. 
\end{enumerate}

\subsection{Training \& Testing}

We use the minibatch Stochastic Gradient Descent (SGD) algorithm coupled with Adagrad \cite{ada} and a learning rate of 0.05 to train each model. Batch size of 32 was used, the embeddings for tags were kept to length 25 and that of songs to 50. The length of the middle layer, $C'$ was kept to 50 and that of the output, $O$ was equal to the number of songs in our dataset, 386046. Dropout regularization with a 0.1 discard probability was used for both the middle and the output layers. We trained the model on a single GTX 1080Ti GPU and the proposed model was implemented using PyTorch\cite{pytorch}.

For testing the models, we adopt the same methodology as followed by Gupta~\cite{own_work}. We iterate through the test histories of the users predicting the next song in the history while giving songs till that point of time as an input to the system. We report $HitRatio@k$~\cite{hit_ratio} where $k$ is the number of songs in the predicted set. We tested two systems based on the attentive neural networks. One was only with the component which takes only the songs into account and not the tags and is referred as SABR(Song Attention Based Recommendation), and the second one with both the components and is referred as STABR(Song and Tag Attention Based Recommendation).

\section{Results and Conclusions}

The results are shown in table 1. Attentive neural networks perform significantly better than all other baseline models and even for Attentive neural networks, the one with the tag component gives a huge gain over the one not having the tag component. This shows that the tags indeed are powerful in modeling the short term user preference and probably the neural network learns a better similarity function than the one proposed by Gupta and hence the gain. 



\end{document}